\documentclass[a4paper,twocolumn]{emulateapj}
\usepackage{amstext}
\usepackage{amsmath}
\usepackage{amssymb}
\usepackage{xcolor,xspace}
\usepackage{multirow,threeparttable,tabularx}
\usepackage[colorlinks=true,linkcolor=blue,citecolor=blue]{hyperref}
\usepackage[super]{nth}
\usepackage{ftnxtra}
\usepackage{aas_macros}

\begin{document}

\shorttitle{SETI via Leakage from Light Sails}

\shortauthors{Guillochon and Loeb}

\title{SETI via Leakage from Light Sails in Exoplanetary Systems}

\author{James Guillochon\altaffilmark{1,2} and Abraham Loeb\altaffilmark{1}}
\altaffiltext{1}{Harvard-Smithsonian Center for Astrophysics, The Institute for Theory and
Computation, 60 Garden Street, Cambridge, MA 02138, USA}
\altaffiltext{2}{Einstein Fellow}

\email{jguillochon@cfa.harvard.edu; aloeb@cfa.harvard.edu}

\begin{abstract} 
The primary challenge of rocket propulsion is the burden of needing to accelerate the spacecraft's own fuel, resulting in only a logarithmic gain in maximum speed as propellant is added to the spacecraft. Light sails offer an attractive alternative in which fuel is not carried by the spacecraft, with acceleration being provided by an external source of light. By artificially illuminating the spacecraft with beamed radiation, speeds are only limited by the area of the sail, heat resistance of its material, and power use of the accelerating apparatus. In this paper, we show that leakage from a light sail propulsion apparatus in operation around a solar system analogue would be detectable. To demonstrate this, we model the launch and arrival of a microwave beam-driven light sail constructed for transit between planets in orbit around a single star, and find an optimal beam frequency on the order of tens of GHz. Leakage from these beams yields transients with flux densities of Jy and durations of tens of seconds at 100~pc. Because most travel within a planetary system would be conducted between the habitable worlds within that system, multiply-transiting exoplanetary systems offer the greatest chance of detection, especially when the planets are in projected conjunction as viewed from Earth. If interplanetary travel via beam-driven light sails is commonly employed in our galaxy, this activity could be revealed by radio follow-up of nearby transiting exoplanetary systems. The expected signal properties define a new strategy in the search for extraterrestrial intelligence (SETI).
\end{abstract}


\section{Introduction}
\begin{figure*}
\centering\includegraphics[width=0.55\linewidth,clip=true]{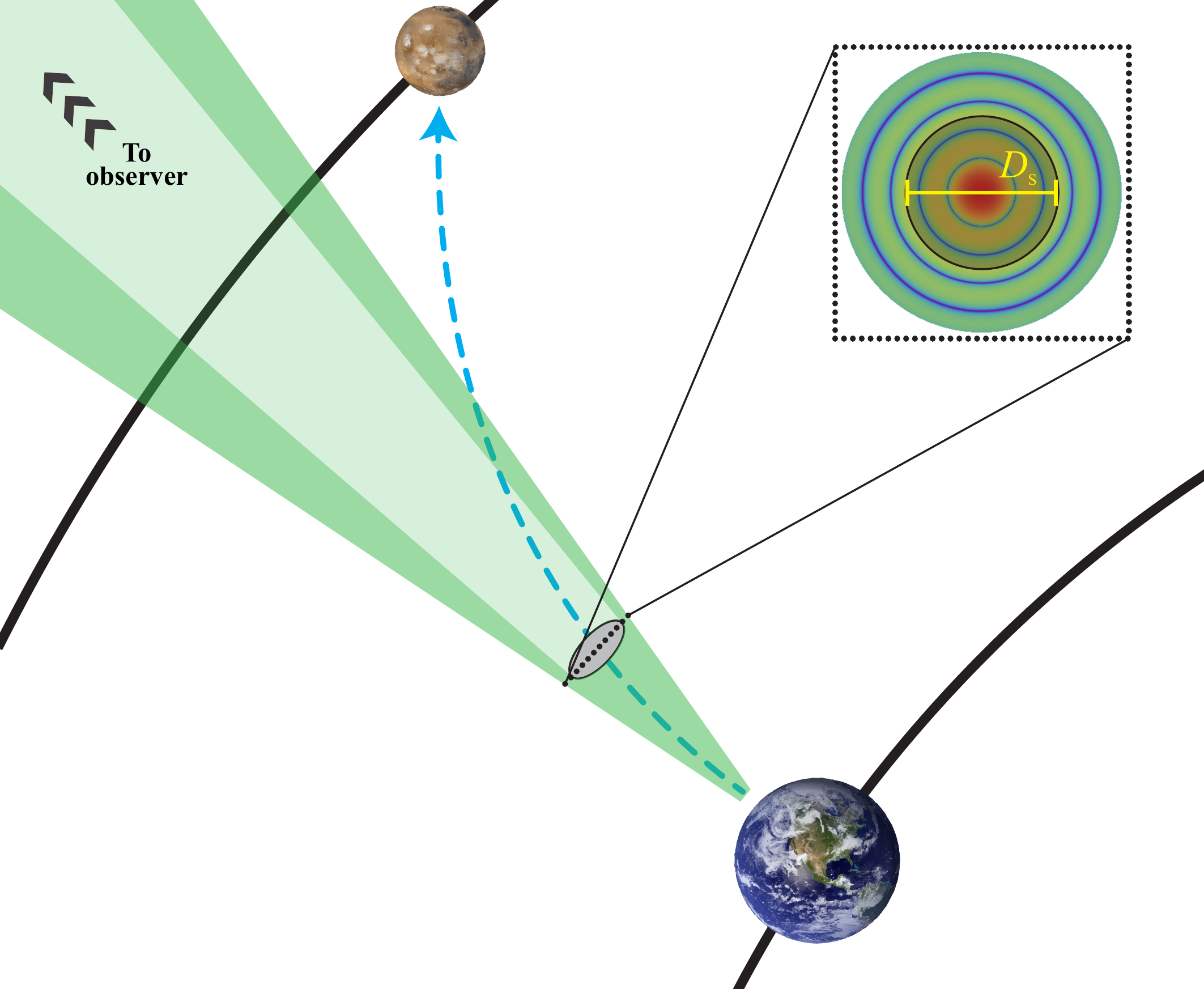}~~~~\includegraphics[width=0.35\linewidth,clip=true]{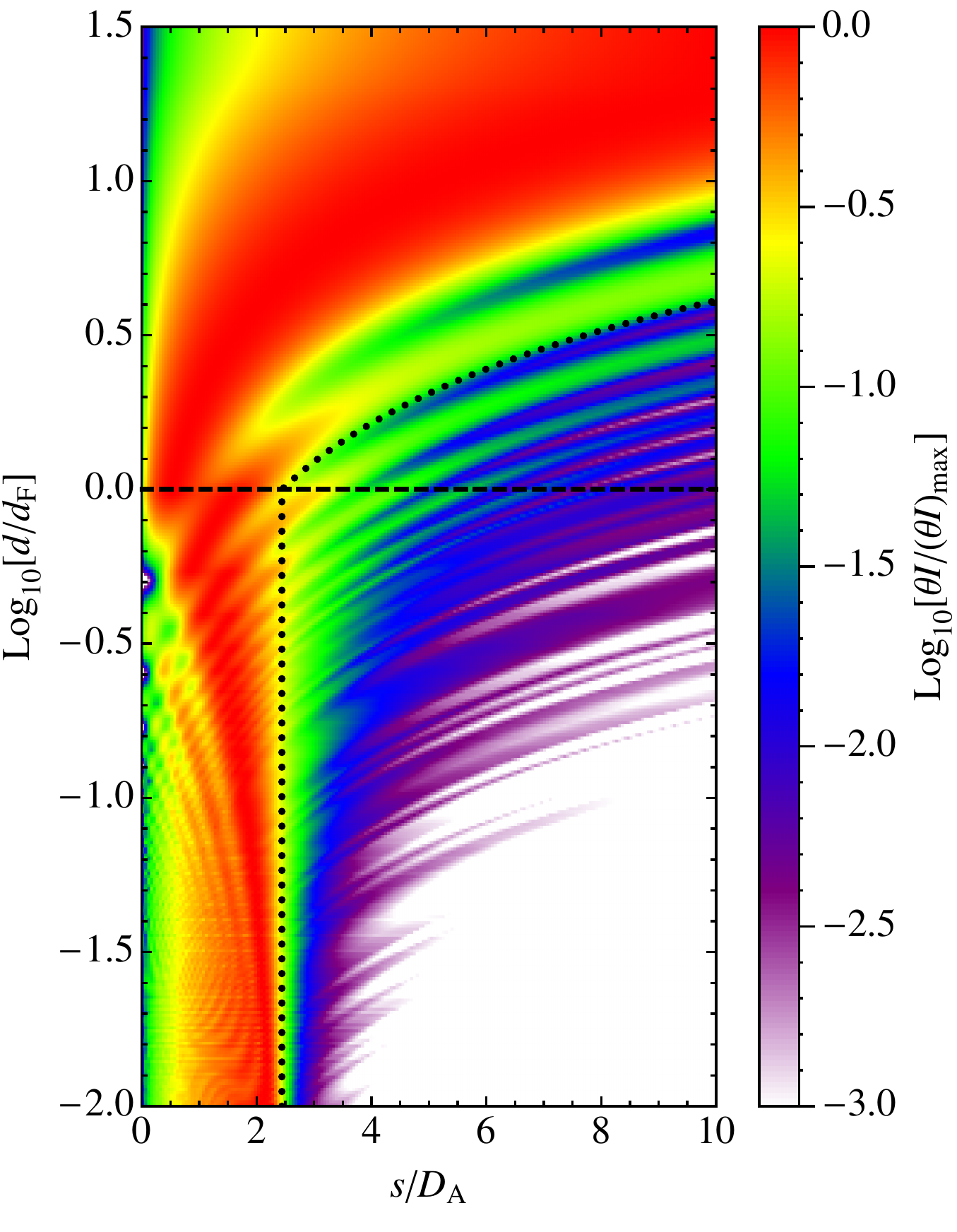}
\caption{{\it Left panel}: Diagram showing the leakage likely for a light sail system developed for Earth-Mars transit. The path of the light sail is shown by the dashed cyan arrow, whereas the beam profile is shown by the green area. The inset shows the logarithm of the intensity $\log I$ within the beam profile, which we have presumed to be in the Fraunhofer regime \citep{Born:1999a,Kulkarni:2014a}. {\it Right panel}: Cylindrical profile of the intensity pattern incident upon the sail, where the cylindrical radius $s$ is scaled to $D_{\rm A}$ and the radial distance $d$ is scaled to $d_{\rm F}$. The dashed line shows $d = d_{\rm F}$, whereas the dotted line shows the region within which most of the intensity is directed.}
\label{fig:diagram}
\end{figure*}

The visitation of Mars by humans by the 2030s is one of the stated goals of NASA\footnote{\url{https://goo.gl/iim1vi}}. Travel time to Mars via chemical rockets is long ($\sim 2$ yr), and thus permanent colonization by humans offers significant advantages over short exploration missions. While many goods will be sourceable on Mars, acquiring them in-situ may be difficult or impossible, and for some goods expediency is particularly valuable \citep{Meyer:1985a}. A practical means of powering such resupply missions are unmanned spacecraft that are propelled via beam-driven light sails \citep{Long:2011a}. In this paper, we construct a mock light sail mission between the Earth and Mars as an analogue to systems in operation around other stars, and consider the detectability of the leakage of radiation from them (Figure \ref{fig:diagram}). While detection of such emission has been suggested previously in the literature \citep{Benford:2010b}, our work represents the first attempt to characterize this emission quantitatively and explore its implications for the search for extraterrestrial intelligence (SETI).

Other than deliberate attempts to contact other civilizations \citep{Horowitz:1993a,Howard:2004a,Benford:2010b,Benford:2010a}, there are seemingly few practical reasons to beam significant amounts of radiation into space. One of the preferred strategies of SETI is to ``eavesdrop'' on communications that may be transmitted between worlds \citep{Cocconi:1959a,Drake:1961a,Loeb:2007a,Tellis:2015a}. While this activity would almost certainly involve beamed radiation, the power requirements of such communication between the planets of a planetary system is relatively small, and thus detection of these transmissions may be difficult. The driving of light sails by artificial means is a practical reason to beam significant amounts of radiation which could be observed. Given its practical appeal for transit within our own solar system, it seems reasonable that intelligent life elsewhere in the galaxy may employ similar technology to facilitate rapid transit between the habitable worlds within its host planetary system. Unlike communications, propelling light sails of any significant mass requires a substantial expenditure of energy, with the isotropic equivalent luminosities of these systems being a non-negligible fraction of their host stars. Such activity could potentially be much easier to detect.

\citet{Benford:2013a} performed a cost-minimization analysis for the construction of such a system, and concluded that a fully-developed system based on existing or near-term technology \citep[the ``interstellar precursor'' column detailed in Table 1 of][]{Benford:2013a} could be constructed for an estimated \$30~B, a cost comparable to a single manned mission to Mars. Such a system would accelerate small spacecraft to 60~km~s$^{\smash{-1}}$, roughly triple the speed of the fastest chemical rockets today, and would be exceedingly cheap to operate once constructed, with electrical costs estimated to be \$40~M per mission. We base much of the design of a mock beamed light sail system on \citet{Benford:2013a}, and argue that such a system would be most energy efficient when the sail is accelerated in the Fresnel regime of the microwave array producing the beamed emission.

In Section \ref{sec:mock} we describe our mock system for rapid Earth-Mars transport via beamed microwave arrays. In Section \ref{sec:leakage} we describe the amount of leakage expected from such a system, and the character of the resulting radio transient. In Section \ref{sec:strategy} we discuss a strategy for optimal expenditure of radio telescope time to maximize the probability of detection. We conclude in Section \ref{sec:conclusion}.

\section{A Beamed Microwave Array for Rapid Earth-Mars Transit}\label{sec:mock}
We have constructed a simple model demonstrating how such a system might operate for mock launch of a 1 metric ton (1000~kg) spacecraft from Earth to Mars. We presume that the spacecraft and its sail are launched from Earth's surface using conventional methods. Once in orbit, a microwave array positioned on Earth's surface beams radiation upon the sail, accelerating the spacecraft to the desired velocity.

Our work closely follows the quantitative assumptions of \citet*{Benford:2013a}, and assumes that the system chosen for Earth-Mars transit would attempt to minimize costs, and should reproduce the qualitative aspects of any practical transit system that utilizes artificially beamed radiation. \citet*{Benford:2013a} suggests that the costs are similar between radiation frequencies of 1~GHz (magnetrons) to 100~Thz (optical lasers), with the costs associated with present-day technology favoring lower-frequency options.

The angular resolution of the accelerating apparatus $\theta_{\rm A} = \lambda/D_{\rm A}$, where $\lambda$ is the wavelength of radiation and $D_{\rm A}$ is the array diameter, was considered by \citet*{Benford:2013a} to determine the pointing accuracy required to drive the sail. However, the angular resolution only describes the beam spread beyond the Fresnel zone in the far-field (i.e. Fraunhofer) regime, i.e. distance $d > d_{\rm F}$, where $d_{\rm F} = D_{\rm A}^{2}/\lambda$ is the Fresnel length \citep{Born:1999a}. In the Fresnel zone, the beam's energy is spread over an angle $\theta_{\rm F} = \max(d_{\rm F}/d,1) \theta_{\rm A}$, which can be significantly larger than the angular resolution. This suggests three design choices to minimize leakage: (i) the Fresnel length should be equal to the distance from Earth at the end of the acceleration phase, $d_{\max} = d_{\rm F}$, (ii) $D_{\rm A}$ and sail diameter $D_{\rm s}$ should be similar (we find $D_{\rm A} \simeq D_{\rm s} / 5$ is optimal), and (iii) the array should use a narrow band centered about single frequency, as different frequencies will have different Fresnel lengths.


Once a spacecraft is in Earth orbit, the required change in orbital energy needs to be at least equal to the difference in the two planets' specific binding energy to their parent star, $\Delta E > G M_{\ast} \left|1 / a_{1} - 1 / a_{2}\right|$, where $M_{\ast}$ is the star's mass and $a_{1}$ and $a_{2}$ refer to the semimajor axes of the two planets. For the case of Earth and Mars this corresponds to a change in velocity of 25~km~s$^{-1}$. However, a direct journey using this minimum amount of energy takes a long time, higher speeds are possible so long as acceleration is maintained over the Fresnel distance. If we employ this strategy, the maximum speed achievable $v_{\rm max} = \sqrt{2 a_{\max} d_{\rm F}}$, and as comfortable travel would not likely exceed an acceleration $a_{\max} \sim 1$~gee, higher speeds are only possible by maximizing $d_{\rm F}$. This suggests that a larger aperture and a shorter wavelength are preferable for the accelerating array.

Fixing $v_{\rm max}$, $D_{\rm A}$, and $a_{\max}$ to fiducial values we find,
\begin{align}
\nu = 68 &\left(\frac{v_{\max}}{100 {\rm~km~s}^{-1}}\right)^2 \left(\frac{D_{\rm A}}{1.5~{\rm km}}\right)^{-2}\nonumber\\
&\times \left(\frac{a_{\max}}{1~{\rm gee}}\right)^{-1}~{\rm GHz}\label{eq:nu},
\end{align}
a frequency compatible with radio surveys such as Parkes and GBT. Equation (\ref{eq:nu}) suggests that microwave frequencies are natural for sailcraft with these parameters, with optical beams being better suited for lower accelerations, smaller spacecraft, and/or higher velocities. The total power of a beam driving a sail via photon momentum (with perfect reflectivity, $\eta = 1$) is
\begin{align}
P_{\rm A} = 1.5 \left(\frac{m_{\rm s}}{1~{\rm metric~ton}}\right) \left(\frac{a_{\max}}{1~{\rm gee}}\right)~{\rm TW},\label{eq:pow}
\end{align}
roughly 10\% of current worldwide consumption and $10^{-5}$ the total solar irradiance of Earth. By comparison, the Saturn V rockets used for the manned lunar missions released 0.1~TW at the time of launch. Using the typical values above, the intrinsic beam-width at $d_{\rm F}$ is $\theta_{\rm A} = 4~\times~10^{-6}$~rad (0.8~arcsec).

A complication of an arbitrarily large launch velocity is that the spacecraft must be slowed down in order to be captured upon reaching its destination. For this reason, our mock model assumes that two microwave arrays exist: one on Earth, and one on Mars, with otherwise identical characteristics. Building two arrays, one on each world, not only permits transit between the two planets at higher speeds, but also enables return trips. Practically, the array on Earth would be built first, which would prolong missions in order to approach Mars at a reasonable speed without the ability to decelerate, but a long-term presence on Mars would certainly benefit from a secondary array constructed there, and presumably space-faring civilizations would build arrays on all worlds between which rapid transit is desired.

\begin{figure}[t!]
\centering\includegraphics[width=0.8\linewidth,clip=true]{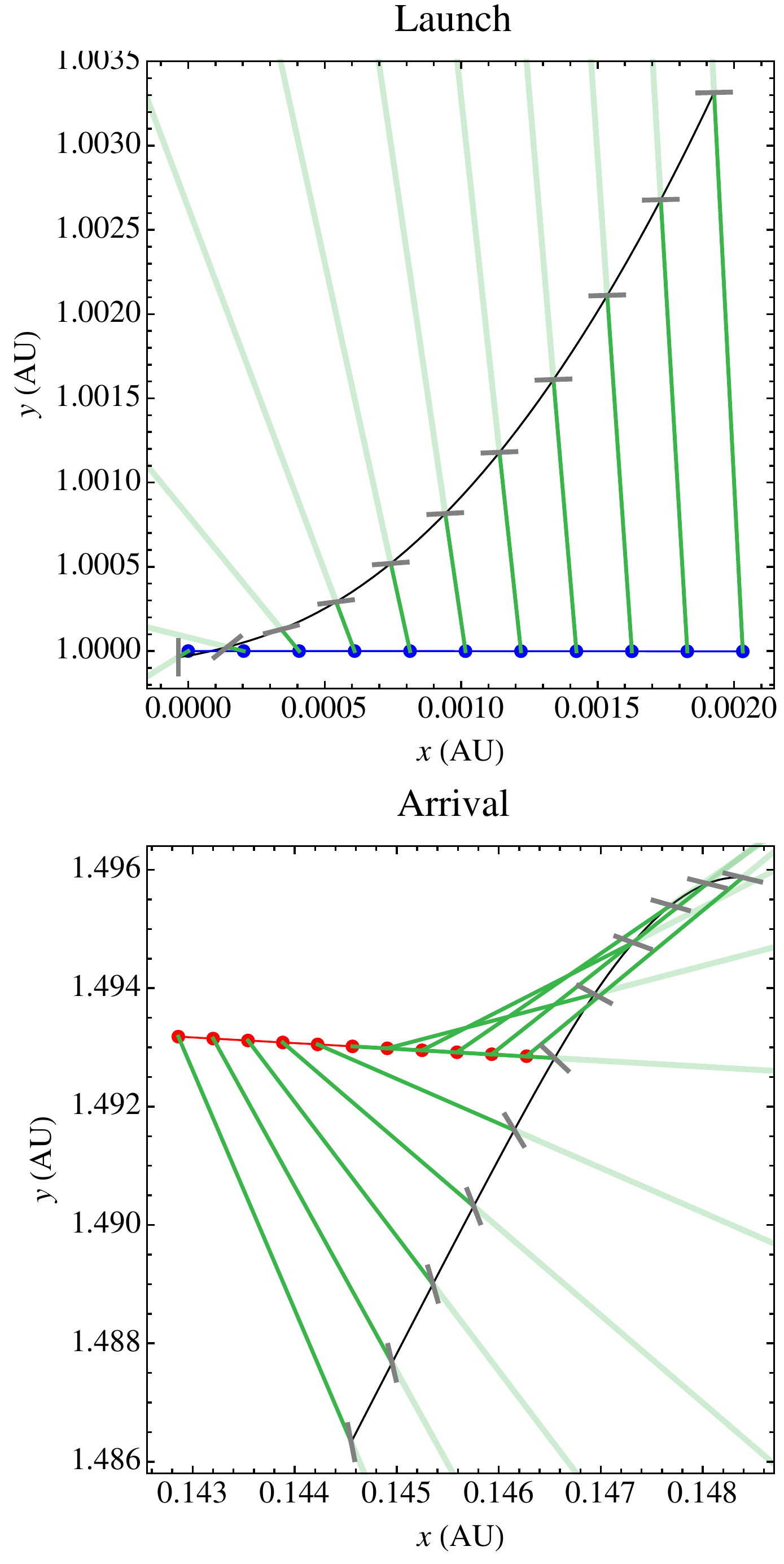}
\caption{A mock launch and arrival of a light sail mission conducted between the Earth and Mars. The top (bottom) panel shows the light sail's launch (arrival), with the black curve showing the path of the sail as it leaves Earth (Mars), represented by the blue (red) curve. The blue (red) points represent Earth's (Mars') position at ten evenly-spaced intervals. The beam direction is shown by the green lines originating from the planets, which impinges upon the sails shown by the gray segments (not to scale), with the angle of the segments showing the angle of the light sail.}
\label{fig:sequence}
\end{figure}

Our model for Earth-Mars transit was constructed in {\tt Mathematica} and is depicted in Figure \ref{fig:sequence}. We account for the Sun, Earth, and Mars' gravity within the model, although we simplify the system somewhat by making the orbits of Earth and Mars circular and coplanar. The light sail is presumed to be constructed of material with a surface density of $4 \times 10^{-6}$~g~cm$^{-2}$ \citep{Matloff:2006a} and carries a payload equal to the sail mass resulting in a total spacecraft mass of 1 metric ton. All other beam/sail properties are set using the fiducial values of Equations~(\ref{eq:nu}) and (\ref{eq:pow}). The strategy for launch is simply to beam in the direction of the line connecting the center of Earth to the sail, ceasing after 3 hours of acceleration (Figure \ref{fig:sequence}, left panel). Deceleration focuses on eliminating the radial velocity component of the spacecraft once it is within Mars' Hill sphere, this is achieved by angling the sail such that the recoil force is oriented in the radial direction ($F \propto \cos^{2} \theta$). After deceleration, the spacecraft is captured into an elliptical orbit about Mars with a periapse distance comparable to Mars' radius.

\section{Light Sail Detection Prospects for Transiting Exoplanetary Systems}\label{sec:leakage}
A fundamental quality of any beamed-radiation propulsion system is the inevitable ``leakage'' of incident radiation around the sides of the spacecraft \citep{Benford:2010b}, as depicted in the left panel of Figure \ref{fig:diagram}. While the beam size would likely be tuned to the sail size to maximize the acceleration while simultaneously minimizing sail mass, the beam size will be difficult to alter during the acceleration/deceleration phases unless the individual antennae are allowed to move, meaning that while the sail may intercept a large fraction of the radiation at early times while it is close to the array, the fraction may decrease as the spacecraft leaves the array's vicinity.

Na\"{i}vely one might expect that the amount of flux intercepted should fall as $d^{-2}$ as the sail moves away from the accelerating array. However, as described in Section \ref{sec:mock}, this is only true in the far-field regime where the distance is greater than the Fresnel distance $d_{\rm F}$; interior to this distance the intensity does not fall off at all (right panel of Figure \ref{fig:diagram}), indeed the energy is beamed through a constant cross-sectional area, rather than a constant solid angle. Because of the already large energy requirements, the most prudent choice is to accelerate the spacecraft interior to this distance to maximize the useful energy. As can be seen in the right panel of Figure \ref{fig:diagram}, the amount of leakage is on order 10\% of the total beamed intensity, with slightly more leakage as $d$ approaches $d_{\rm F}$. Beyond $d_{\rm F}$, leakage grows tremendously, with roughly half of the beam energy being wasted at $d = 3 d_{\rm F}$; this strongly argues for acceleration/deceleration only when $d < d_{\rm F}$.

\begin{figure}
\centering\includegraphics[width=0.8\linewidth,clip=true]{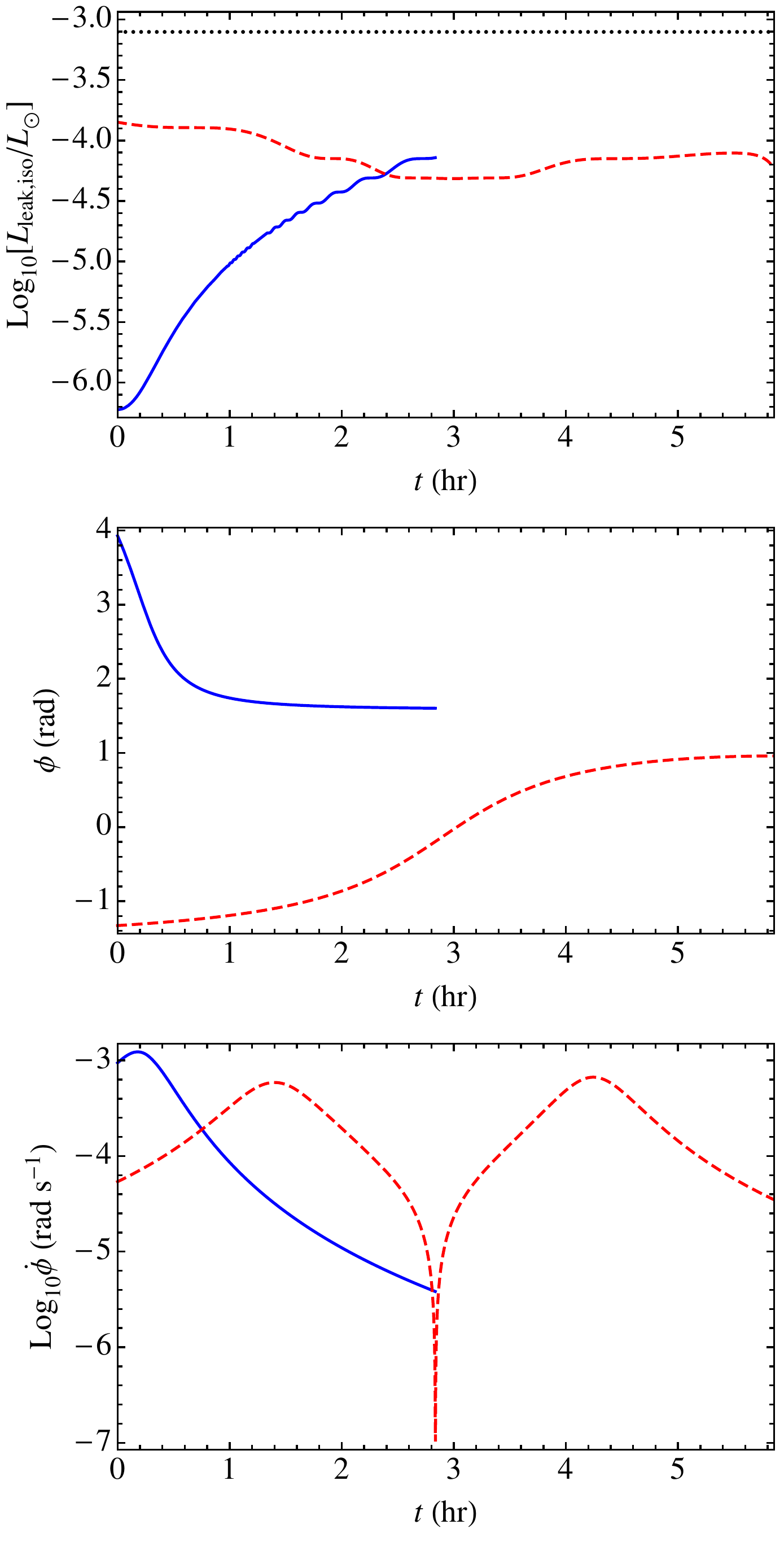}
\caption{Evolution of various quantities over the mock journey from Earth to Mars using a beamed light sail system. Each panel shows how a quantity evolves at launch (blue solid curve) and arrival (red dashed curve). The top panel shows the isotropic equivalent luminosity of leakage from a beam-driven light sail $L_{\rm leak,iso}$ relative to the Sun's total luminosity $L_{\odot}$, with the dotted black line showing the isotropic equivalent luminosity if 100\% of the beam radiation missed the sail, the middle panel shows how the azimuthal angle $\phi$ evolves, and the bottom panel shows its derivative $\dot{\phi}$.}
\label{fig:launch-arrive}
\end{figure}

\begin{figure}
\centering\includegraphics[width=0.9\linewidth,clip=true]{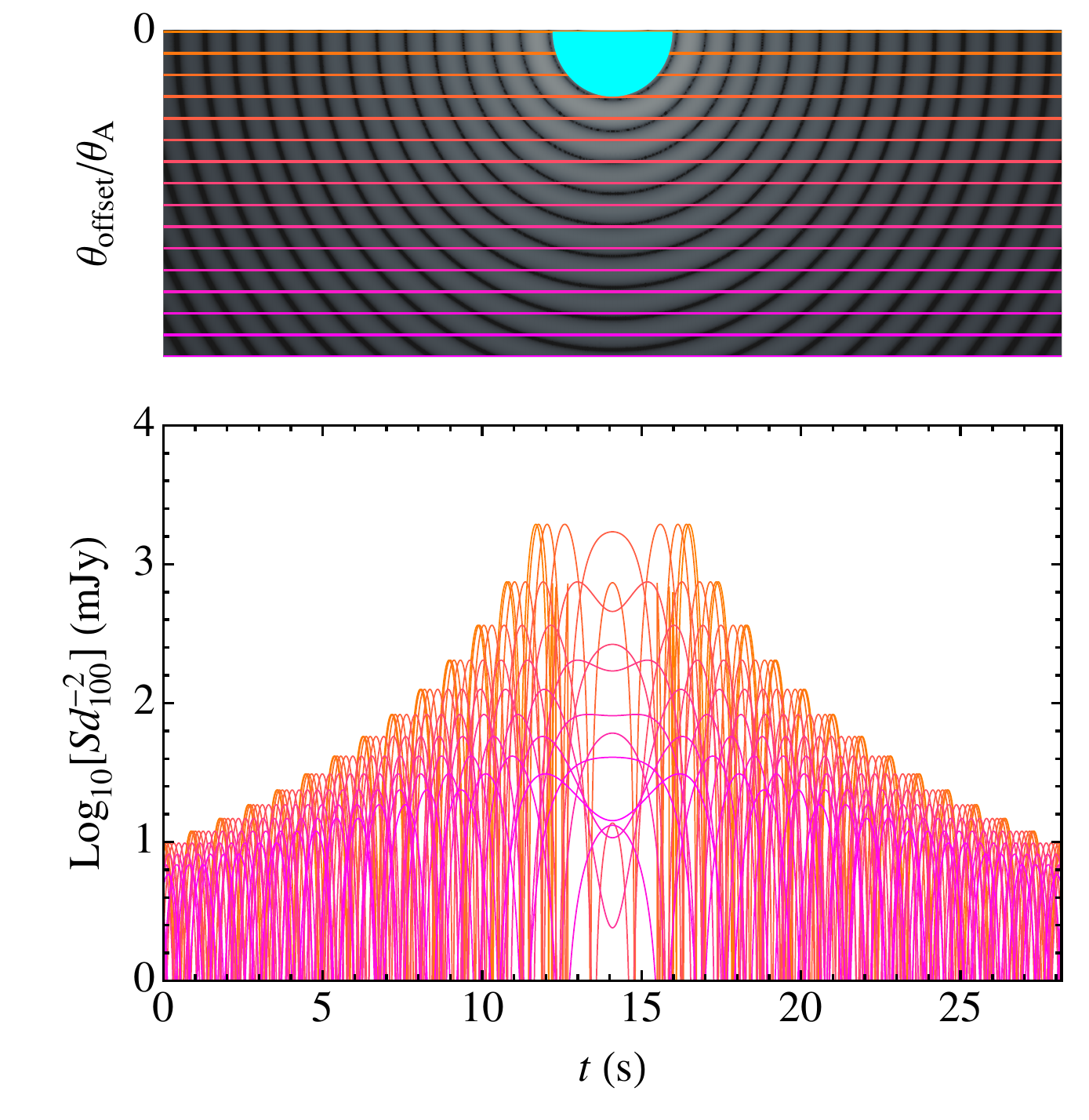}
\caption{Flux density incident upon Earth from a light sail at $d_{100} \equiv d/100~{\rm pc}$ near the end of its acceleration phase where $d = d_{\rm F}$. The grayscale shading in the top panel shows $\log I$, with the horizontal chords tracing the path of the microwave beam at different offset angles $\theta_{\rm offset}$. The orange chord is perfectly centered on the beam ($\theta_{\rm offset} = 0$), whereas the magenta chord represents a grazing event ($\theta_{\rm offset}/\theta_{\rm A} = 10$), the cyan disk shows the region that remains obstructed by the sail. The bottom panel shows the flux density $S$ from each of the matching color chords shown in the top panel as a function of time.}
\label{fig:S}
\end{figure}

Figure \ref{fig:launch-arrive} shows the properties of the beam leakage for the mock mission described in Section \ref{sec:mock}. Because the beam width is so small, its solid angle coverage is only one trillionth the total sky, resulting in isotropic equivalent luminosities that are a factor of $\sim 10^{6}$ times larger than the total radio output of the Sun.

If emitted by a distant civilization in another star system, the peak flux incident upon Earth when $d = d_{\rm F}$ would be
\begin{align}
S_{\rm peak} = 5 &\left(\frac{f_{\rm leak}}{0.1}\right) \left(\frac{\Delta \nu/\nu}{0.1}\right)^{-1} \left(\frac{v_{\max}}{100 {\rm~km~s}^{-1}}\right)^2 \nonumber\\
&\times\left(\frac{r}{100~{\rm pc}}\right)^{-2} \left(\frac{a_{\max}}{1~{\rm gee}}\right)^{-1}~{\rm Jy},
\end{align}
where $f_{\rm leak}$ is the leakage fraction, $\Delta \nu$ is the bandwidth of the array, $r$ is the distance of the system from Earth, and where we have assumed the array bandwidth is equal to the receiver bandwidth.

If the beam were pointed at a single spot while it was active, the small solid angle would make detection difficult. However, because the spacecraft is initially in orbit about its origin body, the beam must track across the sky in order to accelerate it. The middle panel of Figure \ref{fig:launch-arrive} shows that this tracking is quite significant, with the beam sweeping over several radians during both the acceleration and deceleration phases. This sweeping action means that the probability of detection is improved by a factor $\theta_{\rm A}^{-1}$, $\sim 10^{6}$ in our example. Because the beam is only powered for several hours while it sweeps over this large angle (bottom panel of Figure \ref{fig:launch-arrive}), the beam only spends on order seconds pointed in any given direction.

While the spacecraft lies within the array's near-field while it is accelerating, an observer viewing the sail from a distance will record a beam pattern in the far-field regime, with a missing central region obscured by the sail's shadow (Figure \ref{fig:S}). For our example, an observer will record a series of short bursts a few tenths of a second in duration as the beam's path cuts a chord across the observer's line of sight. For chords that cut through the beam pattern's center, the duration of each burst will be roughly constant (orange lines in Figure \ref{fig:S}), but off-center crossing events, which are more common, will exhibit bursts with initially short durations that lengthen as the chord cuts across the beam pattern's axis of symmetry. In either case, a symmetrical transient is produced with a time-variable intensity.

\begin{figure}
\centering\includegraphics[width=0.9\linewidth,clip=true]{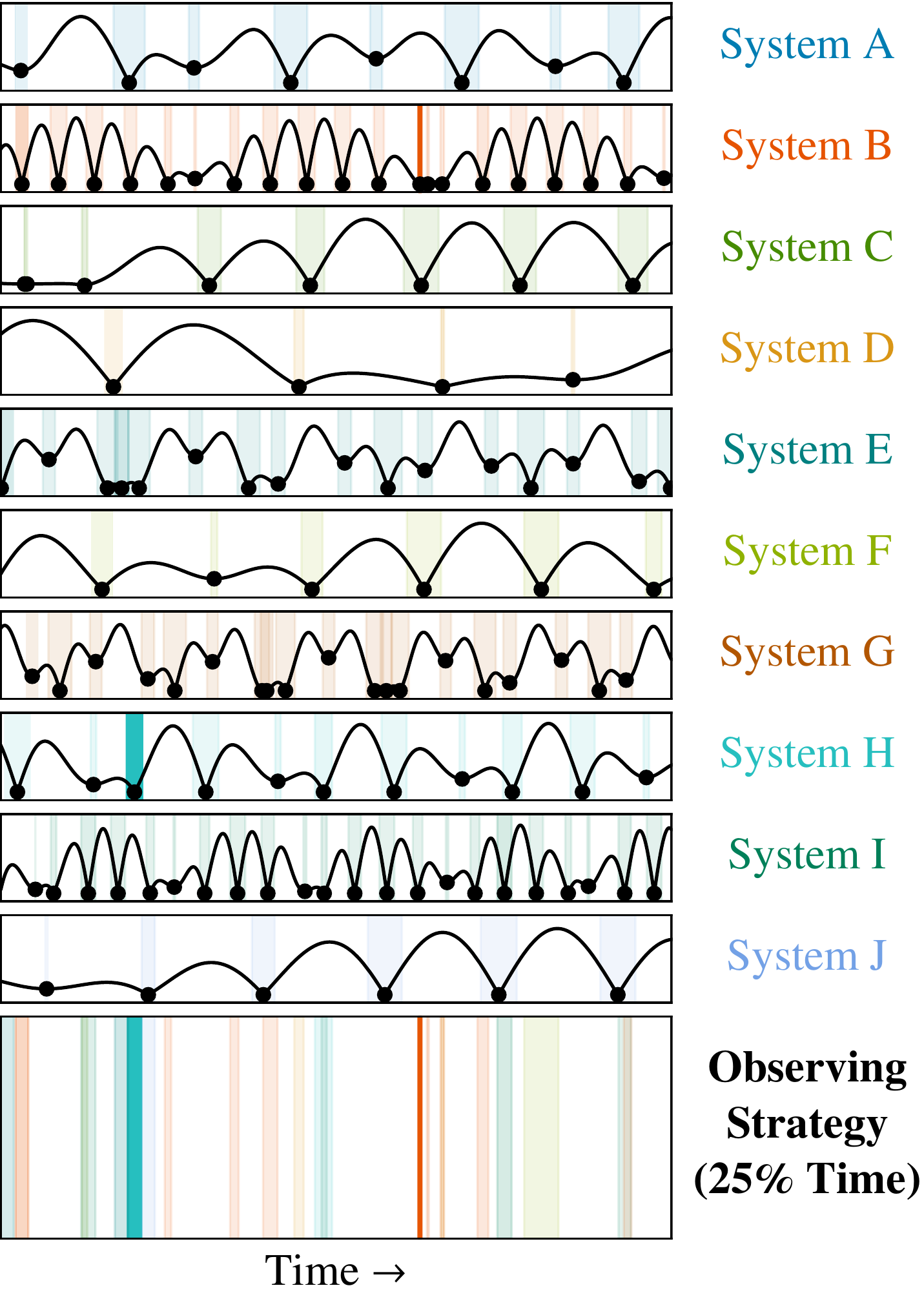}
\caption{Optimal times to search for light sail leakage from simulated sample of multiply-transiting systems. The first ten panels (labeled System A -- J) show the projected distance in the image plane between each pair of planets as a function of time (black curves), with the black points denoting the local minima of these functions, periods when the beam is more likely to be facing the observer. We assume that the period of enhanced likelihood is comparable to the travel time between the two planets (where we use Earth-Mars analogues and have set $v = 100$~km~s$^{\smash{-1}}$), as indicated by the colored bands centered about each point. The opacity of each colored band represents the priority of a given event, which is taken to the be the inverse of the product of the distance in the projection and the travel time. The bottom panel shows the top priority events to follow up (color coded to match the individual pairs in the top ten panels), with the restriction that only 25\% of the telescope's time will be available for SETI.}
\label{fig:strategy}
\end{figure}

\section{Optimal Strategy for Detection of Leakage from Beamed Light Sails}\label{sec:strategy}

The transient signal presented in the previous section is bright enough to be detectable even when $\theta \sim 10 \theta_{\rm A}$ with most SETI radio searches ($S \sim 10$~mJy), yielding a detection probability $P_{\rm det} \sim 10 \theta_{\rm A} \sim 10^{-4}$. If such a system were constructed for Earth-Mars transit, the minimum time between missions would be the acceleration time of the spacecraft; in our example this is a couple of hours. As the infrastructure has a high capital cost but low operating expenses, it seems reasonable to expect that the apparatus would be in constant use by any civilization that has built it. If this is the case, then the only obstacle is geometric: the spacecraft needs to be accelerated along a path that lies along our line of sight. This suggests that the probability for detection is maximized when the two planets are in conjunction with one another (in projection) as observed from Earth \citep{Siemion:2014a,Siemion:2015a}, with the probability of detection being maximal for systems in which the conjunction is as small as possible.

Because transiting exoplanetary systems often provide all six Kepler elements describing the orbits of each transiting planet, the projected conjunctions for such systems can be predicted exquisitely. Additionally, such systems are guaranteed to have projected vertical separations that are no greater than the stellar radius $R_{\ast}$, which increases the probability of detection by a factor $a/R_{\ast}$, where $a$ is the planet's semimajor axis (a few hundred for an Earth-Mars analogue). Note however that this does {\it not} necessarily mean that the two planets need to be transiting their host star at the same time, only that the two planets are in close proximity to one another in projection. In fact, closer projected conjunctions are likely to occur when the planets are not transiting their parent star if they possess two different inclinations.

This suggests an observing strategy in which systems are observed near projected conjunctions for a timescale comparable to the travel time between the two planets in a given system, with preference given to the closest projected conjunctions, and for conjunctions in which the travel time is at a minimum (i.e. true conjunctions). Figure \ref{fig:strategy} shows an optimal strategy for detecting the light sail leakage transients described here given a hypothetical list of multiply-transiting systems and the 25\% observing time allocated on the Parkes and GBT radio telescopes by the {\it Breakthrough Listen} initiative\footnote{\url{http://www.breakthroughinitiatives.org}}. From this figure, it is clear that certain systems should have higher priority than others, but that multiple targets should be observed to improve the odds of detection. If monitoring were continuous, the beam size of $\sim 10^{-4}$~rad multiplied by $a/R_{\ast}$ suggests a probability of detection on the order of $1\%$ per conjunction. For a five year survey duration with $\sim 10$ conjunctions per system, this implies $\sim 10$ multiply-transiting, inhabited systems would need to be tracked to guarantee a detection for an apparatus similar to what is detailed here.

\section{Conclusion}\label{sec:conclusion}
In this Letter, we have presented a mock beam-driven light sail system enabling the transport of spacecraft between the Earth and Mars, and have calculated the observability of analogues to such apparatuses as seen by a distant observer. The particular example design we have presented here is chosen on the basis of optimizing energy use and minimizing cost. We have shown that there is a single optimal frequency for such apparatuses given a desired acceleration, speed, and sail size (Equation \ref{eq:nu}) which falls into the microwave band, producing transients with $S \sim$~few Jansky at 100 pc and durations of tens of seconds, with a probability of detection on the order of 10\% per inhabited system for a five year survey. We have ignored a number of engineering challenges in our mock transit system, the solutions to which may affect the observable transient signals. But because our system is motivated by minimizing wasted energy, we reason that any beam-driven light sail system is likely to employ a strategy similar to what is presented here.

\bigskip
\acknowledgements
We thank J.~Benford, J.~Wright, and the anonymous referee for comments. This work was supported by Einstein grant PF3-140108 (J.~G.).
\bigskip\bigskip

\bibliographystyle{apj}

\end{document}